\documentclass[12pt]{article}
\usepackage{amsfonts}
\usepackage{enumerate}
\newcommand{\mId}{\mbox{\rm id}}
\newcommand{\mR}{{\mathbb R}}

\newcommand{\arctanh}{\mbox{\rm arctanh}}
\author{Cezary Gonera\\
Dept. of Theoretical Physics II\\
Pomorska 149/153, 90-236 \L{}\'od\'z,Poland}
\title{More about generalized maximally superintegrable systems of Winternitz type}
\begin{document}
\maketitle
\abstract{\scriptsize
Recently proposed procedure of constructing maximally superintegrable systems of 
Winternitz type is further developed and illustrated by an example of a system admitting 
an explicit construction of angle variables  and additional integrals of motion.

A possible application of the method to Liouville system is briefly presented.

\noindent Pacs numbers 02.60Lj
}
\vspace{0.5cm}

Recently we have outlined  \cite{1} 
general procedure leading to superintegrable models generalizing Winternitz 
system \cite{1}--\cite{6}. 
We have shown that for completely separated hamiltonians the potentials corresponding to 
superintegrable models have the form $u(x)=\beta^2(\phi(x)-x)^2$ with $\phi(x)$ having property 
$\phi(\phi(x))=x$. 

The system consisting of the set of the independent degrees of freedom, each
governed by the natural hamiltonian, with the potential energy of the above form  and the mass $\mu_k$ is
maximally superintegrable iff all ratios $\frac{\beta_k}{\sqrt{\mu_k}}/\frac{\beta_l}{\sqrt{\mu_l}}$
are rational.

In the present letter we give further examples of our method. In particular, we show how it
 can be applied to more general separable theories like Liouville systems.

The key role in our approach is played by the function $\phi(x)$ obeying $\phi\circ\phi=\mId$. 
It is not difficult to classify such $\phi$'s.

Assume that $\phi$ has the following properties:
\begin{enumerate}[i)]
\item $\phi:\mR\rightarrow\mR$ is continuously differentiable and onto
\item $\phi\circ\phi=\mId$
\end{enumerate}
Then $\phi$ is one-to-one, $\phi^{-1}=\phi$, and  $\phi$ is either strictly increasing or strictly 
decreasing. In the former case $\phi(x)=x$ while in the latter
\begin{equation}
\phi(x)=F(c-F^{-1}(x))
\label{w1}
\end{equation}
where $F:\mR\rightarrow\mR$ is continuously differentiable one-to-one mapping of $\mR$ onto $\mR$.

To show this let us first assume that $\phi(x)$  is strictly increasing. Then $\phi(x)\geq x$
implies $x=\phi(\phi(x))\geq\phi(x)$ so that $\phi(x)=x$ and the same applies if $\phi(x)\leq x$.

Assume now that $\phi$ is strictly decreasing. Then $\phi(x)=x$ has exactly one solution, $x=a$.
Put $c=2a$ and define 
\[
F(x)=\left\{
\begin{array}{ll}
x, & x\in(-\infty,a\rangle\\
\phi(c-x), & x\in(a,\infty)
\end{array}
\right.
\]
It is then easy to check that $F:\mR\rightarrow\mR$ is one-to-one continuously differentiable and onto and
eq.(\ref{w1}) holds. Note that the representation (\ref{w1}) is by far nonunique.

Having a general solution~(\ref{w1}) at our disposal we can construct a wide class of 
superintegrable systems following prescription given in~\cite{1}.
The relevant completely separated hamiltonians read
\begin{equation}
\begin{array}{l}
\displaystyle
H=\sum_{k=1}^N(\frac{p_k^2}{2\mu_k}+U_k(x_k))\equiv\sum_{k=1}^NH_k(x_k,p_k)\\
\displaystyle\mbox{\rm where}\\
\displaystyle U_k(x)\equiv\beta_k^2(\phi_k(x)-x)^2,\ \ \beta_k^2=\frac{\alpha^2m_k^2\mu_k^2}{8}
\end{array}
\label{w2}
\end{equation}
and  $m_k$ are arbitrary integers.

To see how it works let us take simple $F$ giving rise to a nontrivial potential:
\begin{equation}
F^{-1}(x)=\varrho x^3
\label{w3}
\end{equation}
where $\varrho$ is a constant of dimension of inverse length squared. Then with $c$ being another 
constant of dimension of length we get:
\begin{equation}
\phi(x)=\sqrt[3]{\displaystyle\frac{c-\varrho x^3}{\varrho}}
\label{w4}
\end{equation}
and ($\sigma_k\equiv c_k/\varrho_k$)
\begin{equation}
H=\sum_{k=1}^N(\frac{p_k^2}{2\mu_k}+\frac{\alpha^2\mu_k^2m_k^2}{8}(\sqrt[3]{\sigma_k-x_k^3}-x_k)^2)
\label{w5}
\end{equation}

The general considerations of \cite{1} give at once the action variables while for obtaining
 the angle variables an explicit integration is necessary which, in our example, is not easy. Once
the angle variables are known it is straightforward (in principle, at least) to find the additional 
integrals of motion (see, for example \cite{7,8}). However, in general it appears that 
the relevant integrations cannot be performed explicitly. Therefore, our construction gives a very
wide class of superintegrable systems with rather complicated, in general, additional integrals of motion.
It means in particular that the freedom in choice of the variables which separate the
Hamilton-Jacobi equation appears only on the level of general canonical transformations 
(the additional integrals are not quadratic in momenta).

To show more explicitly how that  works  consider another simple choice:
\begin{equation}
F(x)=a\sinh\left(\frac{x}{a}\right)
\label{w6}
\end{equation}
Then $\phi(x)$ reads
\begin{equation}
\phi(x)=a(\sinh\left(\frac{c}{a}\right)\sqrt{1+\frac{x^2}{a^2}}-\cosh\left(\frac{c}{a}\right)\frac{x}{a})
\label{w7}
\end{equation}
and the relevant model is ($\sigma_k\equiv\frac{c_k}{a_k}$)
\begin{eqnarray}
H&=&\sum_{k=1}^N\left(\frac{p_k^2}{2\mu_k}+\beta_k^2a^2_k(2(1+\cosh\sigma_k)\cosh\sigma_k
\frac{x_k^2}{a_k^2}-{}\right.\nonumber\\
&&\left.{}-2\sinh\sigma_k(1+\cosh\sigma_k)\frac{x_k}{a_k}\sqrt{1+\frac{x_k^2}{a_k^2}}
+\sinh^2\sigma_k)\right)
\label{w8}
\end{eqnarray}
The advantage of this model is that one can find the angle variables by elementary integration.

In terms of action variables our hamiltonian reads \cite{1}
\begin{equation}
H=\sum_{k=1}^NE_k=\alpha\sum_{k=1}^Nm_kJ_k
\label{w9}
\end{equation}
therefore,
\begin{equation}
\frac{\partial E_k}{\partial J_k}=\alpha m_k.
\label{w10}
\end{equation}
The angle variables are given by
\begin{equation}
\label{w11}
\phi_k=\alpha\mu_km_k\int^{x_k}\frac{\displaystyle dx_k}{\displaystyle \sqrt{2\mu_k(E_k-U_k)}}
\end{equation}
and the lower integration limit may be chosen arbitrarily. In our case the integration can be performed 
explicitly. The result reads:
\begin{eqnarray}
\phi_k&=&\arcsin\left(\frac{\displaystyle 2\beta_ka_k\cosh\left(\frac{\sigma_k}{2}\right)}
{\displaystyle \sqrt{E_k}}\left(\frac{x_k}{a_k}\cosh\left(\frac{\sigma_k}{2}\right)+{}\right.\right.
\nonumber\\
&&\left.\left.{}-\sqrt{1+\frac{x^2_k}{a^2_k}}\sinh\left(\frac{\sigma_k}{2}\right)\right)\right)
+{}\nonumber\\
&&{}+\tanh\left(\frac{\sigma_k}{2}\right)\arcsin\left(
\frac{\displaystyle 2\beta_ka_k\cosh\left(\frac{\sigma_k}{2}\right)}{\displaystyle
\sqrt{E_k+4\beta_k^2a_k^2\cosh^2\left(\frac{\sigma_k}{2}\right)}}\cdot{}\right.\nonumber\\
&&\cdot\left.\left(\sqrt{1+\frac{x^2_k}{a^2_k}}\cosh\left(\frac{\sigma_k}{2}\right)
-\frac{x_k}{a_k}\sinh\left(\frac{\sigma_k}{2}\right)\right)\right)
\label{w12}
\end{eqnarray}
The angle $\phi_k$ is well-defined. Indeed, the first term on the right hand side changes by $2\pi$ 
while the second comes  back to the initial value when $x_k$ attains again initial value
(for that reason $\tanh\left(\frac{\sigma_k}{2}\right)$ can be also arbitrary).

In order to construct the additional integral of motion let us consider the simplest case 
$N=2$, $m_1=m_2=1$,  $\mu_1=\mu_2=\mu$, ,$a_1=a_2=a$, $\sigma_1=\sigma_2=\sigma$ and write (\ref{w12})
as
\begin{equation}
\phi_k = \psi_k+\tanh\left(\frac{\sigma}{2}\right)\chi_k, \ \ \ k=1,2
\label{w13}
\end{equation}
Under the above assumptions $\dot{\phi}_1=\dot{\phi}_2$, so $\phi_1-\phi_2$ is an integral of motion.
Morover, any periodic function of $\phi_1-\phi_2$ is well defined on phase space so that,
for example $\sigma(I_1,I_2)\sin(\phi_1-\phi_2)$ can be taken as an additional integral of motion
(for arbitrary $\sigma(I_1,I_2)$).

Howewer, there is still some technical obstacle. Due to the properties of $\phi_k$ mentioned above
any periodic function of $\phi_k$'s can be expressed in terms $x_k$'s and $p_k$'s;
in general the resulting expression will be very complicated and not expressible in terms
of elementary functions. The exceptional case would be if the periodic functions
 of $\phi_k$ are expressible in terms of trigonometric functions of both $\psi_k$ and $\chi_k$.
This is possible if $\tanh\left(\frac{\sigma}{2}\right)$ is rational.
In particular we assume that $\tanh\left(\frac{\sigma}{2}\right)=\frac{1}{2}$, i.e. 
$\sigma=\ln 3$.
{\bf Let us stress again that this condition is of technical nature and has nothing to do with
the superintegrability condition.}

Now, under our assumptions
\begin{equation}
\sin2(\phi_1-\phi_2)=\sin(2(\psi_1-\psi_2)+(\chi_1-\chi_2))
\label{w14}
\end{equation}
is an integral of motion. It is polynomially expressible in terms of $\sin\psi_k$, $\cos\psi_k$,
$\sin\chi_k$,  $\cos\chi_k$. In turn, 
\begin{equation}
\begin{array}{lcl}
\displaystyle\sin\psi_k & = & \frac{4\beta a}{\sqrt{3H_k}}\left(\frac{2x_k}{\sqrt{3}a}-\frac{1}{\sqrt{3}}\sqrt{
1+\frac{x_k^2}{a^2}}\right)\\
&&\\
\displaystyle\sin\chi_k&=&\frac{4\beta a}{\sqrt{3H_k+16\beta^2a^2}}\left(\frac{2}{\sqrt{3}}\sqrt{1+\frac{x_k^2}{a^2}}
-\frac{1}{\sqrt{3}}\frac{x_k}{a}\right)\nonumber\\
&&\\
\displaystyle\cos\psi_k&=&\frac{p_k}{\sqrt{2\mu H_k}}\\
&&\\
\displaystyle\cos\chi_k&=&\sqrt{\displaystyle\frac{p_k^2+\frac{32}{3}\beta^2a^2\mu}{2\mu H_k}}
\end{array}
\label{w15}
\end{equation}
The sign ambiguity for cosines of $\psi_k$'s is properly taken into account by replacing 
$\sqrt{E_k-U_k(x)}$ by $p_k/\sqrt{2\mu}$. Also the form of $\cos\chi_k$ shows that $\chi_k$
is well defined over Liouville torus.

Let us note that even in this simplest case the additional integral cannot be chosen as a polynomial
function of momenta.

The original Winternitz model was defined on semiaxis. It is not difficult to generalize our
discussion to this case. The counterpart of eq.(\ref{w1}) reads now
\begin{equation}
\phi(x)=F\left(\frac{c}{F^{-1}(x)}\right)
\label{w16}
\end{equation}
where $F:\mR_{+}\rightarrow\mR_{+}$ is continuously differentiable isomorphism. Indeed, 
$l=\ln x$ is a smooth inverible mapping from $\mR_{+}$ to $\mR$; therefore, if $\phi:\mR\rightarrow\mR$
and $\phi\circ\phi=\mId$ then $\tilde{\phi}=l^{-1}\circ\phi\circ l:\mR_{+}\rightarrow\mR_{+}$ obeys
also $\tilde{\phi}\circ\tilde{\phi}=\mId$. Taking $F(x)\sim x^\lambda$ one obtains Winternitz
model. Again a variety of models can be constructed by selecting various $F$. For example,
\begin{equation}
F(x)=x\left(1+\frac{x}{a}\right)
\label{w17}
\end{equation}
gives
\begin{equation}
\phi(x)=
\frac{\displaystyle c\left(\sqrt{1+\frac{4x}{a}}+1\right)}{\displaystyle 2x}
\left(1+
\frac{\displaystyle c\left(\sqrt{1+\frac{4x}{a}}+1\right)}{\displaystyle 2ax}\right)
\label{w18}
\end{equation}
We obtain a deformation of Winternitz model which is recovered in $a\rightarrow\infty$ limit.

Our method is by far not restricted to the hamiltonians of the form given by eq.(\ref{w2}).
Consider the Liouville system (see \cite{9} for example)
\begin{equation}
H=\frac{\displaystyle \sum_{k=1}^N\left(\frac{p_k^2}{2\mu_k}+V_k(x_k)\right)}{\displaystyle
\sum_{k=1}^Nc_k(x_k)}
\label{w19}
\end{equation}
It is separable because the H-J equation $H(x,p)=E$ separates into the set of equations
\begin{equation}
\frac{p_k^2}{2\mu_k}+(V_k(x_k)-Ec_k(x_k))=\epsilon_k, \ \ \ \sum_{k=1}^N\epsilon_k=0
\label{w20}
\end{equation}
corresponding to totally separable hamiltonian
\begin{equation}
\tilde{H}(x,p,E)=\sum_{k=1}^N\left(\frac{p_k^2}{2\mu_k}+V_k(x_k)-Ec_k(x_k)\right)
\label{w21}
\end{equation}
The total energy $E$ enters now as a parameter of effective potentials
\begin{equation}
U_k\equiv V_k(x)-Ec_k(x)
\label{w22}
\end{equation}

We see that our method can be applied if we can select potentials which depend linearily on some parameter
$E$. One straghtforward choice is to take
 $V_k(x_k)=c_k(x_k)=\beta^2_k(x_k-\phi_k(x_k))^2$
as in eq.(\ref{w2}); also in the original Winternitz model one can separate
the potential into two pieces in arbitrary way.
To see that less obvious possibilities exist consider the model given by eq.(\ref{w6}). The 
potential in eq.(\ref{w8}) can be rewritten as 
\begin{equation}
U(x)=(V(x)-Ec(x)+U_0)\gamma(E)
\label{w23}
\end{equation}
where the following identifications have been made
\begin{eqnarray}
V(x)&\equiv&2\beta^2x^2 \label{w24}\\
c(x)&\equiv&\frac{x}{a}\sqrt{1+\frac{x^2}{a^2}}\label{w25}\\
E&=&2\beta^2 a^2\tanh\left(\frac{c}{a}\right)\label{w26}\\
U_0&=&-\frac{1}{2}\sqrt{4\beta^2a^4-E^2}+\beta a^2\label{w27}\\
\gamma(E)&=&
\frac{\displaystyle 1}{\displaystyle \sqrt{1-\frac{E^2}{2\beta^2a^2}}}
\left(
\frac{\displaystyle 1}{\displaystyle \sqrt{1-\frac{E^2}{2\beta^2a^2}}}+1\right)\label{w28}
\end{eqnarray}
Consider now the Liouville model given by
\begin{equation}
H=\frac{\displaystyle \sum_{k=1}^N\left(\frac{p_k^2}{2\mu_k}+2\beta_k^2x_k^2\right)}{
\displaystyle 1+\sum_{k=1}^N\frac{x_k}{a_k}\sqrt{1+\frac{x_k^2}{a_k^2}}}, \ \ \ \ 
\begin{array}{l}
\displaystyle\beta_k^2=\frac{\alpha^2\mu_km_k^2}{8},\\
\\
\displaystyle\beta_1^2a_1^2=\ldots=\beta_N^2a_N^2
\end{array}
\label{w29}
\end{equation}
where $m_k$ are arbitrary integers. For $|E|<\min_k(2\beta_k^2a_k^2)$ the motion is bounded with
respect to all variables $x_k$. Morover, one can choose 
$c_k=a_k\arctanh\left(\frac{E}{2\beta_k^2a_k^2}\right)$ to find that our model is superintegrable.

In fact, the equivalent hamiltonian (\ref{w21}) is given by eq.(\ref{w24}--\ref{w28}) with the replacement 
$\beta_k^2\rightarrow\beta_k^2/\gamma(E)$; note that the function $\gamma(E)$ is
universal, i.e. does not depend on $k$.

As a special example consider the case $m_k=1$; then $\beta_1=\ldots=\beta_N=\beta$, 
$a_1=\ldots a_N=a$ and we obtain the deformation of isotropic oscillator which is still superintegrable.

More general discussion of Liouville and Staeckel systems will be given elsewhere.

\vspace{1cm}
{\noindent\bf\Large Acknowledgements}\linebreak

\noindent Numerous very useful discussions with P. Kosi\'nski and P. Ma\'slanka are
gratefully acknowledged.

\noindent The author is supported by KBN grant, no 5 P03B 060 21.


\begin{thebibliography}{9}

\bibitem{1} 
C. Gonera, P. Kosi\'nski, P. Ma\'slanka, Phys. Lett. A289, 66, (2001)

\bibitem{2}
P. Winternitz et al., Sov. J. Nucl. Phys. 4, 444 (1966)

\bibitem{3}
J. Fris et al., Phys. Lett. 16, 354 (1965)

\bibitem{4}
A. Makarov et al. Nuovo Cimento 52, 1061 (1967)

\bibitem{5} 
N. W. Evans, J. Math. Phys. 32, 3369 (1991)

\bibitem{6} 
N. W. Evans, Phys. Lett. A 147, 483 (1990)

\bibitem{7}
L. Landau, E. Lifshitz, {\em Mechanics}\/, Pergamon Press, 1976 

\bibitem{8}
V. Arnold, {\em Mathematical Methods of Classical Mechanics}, Springer, 1978

\bibitem{9}
A. M. Perelomov, {\em Integrable Systems of Classical Mechanics and Lie Algebras}\/, Birkh\"auser, 1990

\end{thebibliography}
\end{document}